\documentclass[a4paper,11pt]{article}
\usepackage[ams]{}
\usepackage{amsmath}
\usepackage{amsthm}
\usepackage{hyperref}
\usepackage{amssymb}
\usepackage{graphicx}
\usepackage[all,arc]{xy}
\usepackage[theorem]{}

\theoremstyle{plain}			

\theoremstyle{definition}		

\begin{document}
\title{On a Final Theory of Mathematics and Physics}
\author{Michael Rios\footnote{email: michael.rios@bakersfieldcollege.edu}\\\\\emph{Bakersfield College}\\\emph{Department of Mathematics}
  } \date{\today}\maketitle
\begin{abstract}
Since ancient times, mathematics has proven unreasonably effective in its description of physical phenomena.  As humankind enters a period of advancement where the completion of the much coveted theory of quantum gravity is at hand, there is mounting evidence this ultimate theory of physics will also be a unified theory of mathematics.  
\\\\
$Keywords:$ Twistors, M-theory, Gravity, Motives.
\end{abstract}

\newpage
\tableofcontents
\section{Introduction}
Since the days of ancient Babylon, where Pythagorean triples were tabulated, to the proof of Fermat's last theorem in 1994, the interplay between mathematics and physics has been abundantly fruitful.  The Pythagoreans of old believed in harmony between the apeiron and peiron (emptiness and form) where the inhalation of the apeiron by the peiron is responsible for the existence of physical reality itself \cite{1}.  Through a limiting transformation of the umlimited void, number and physical reality in this framework are born together, simultaneously.

At the dawn of an era of quantum gravity, the spirited vision of the Pythagoreans lives on, as theoretical physics is confronted with the problem of constructing spacetime, from a deeper, primordial theory.  Such a theory seems to require the creation of new mathematics, as Newton did to describe motion via his method of fluxions \cite{2}, giving rise to the calculus.  Presently, the most promising candidate for quantum gravity is M-theory, a theory still yet to be completed but has its origins in Veneziano's 1968 dual resonance model \cite{3}.  It was not until 1974 that Scherk and Schwarz demonstrated \cite{4} that the \emph{string theory} emerging from the dual resonance model contains a graviton, hence a quantum description of gravity.  A quantum theory of gravity, in contrast to Einstein's elegant picture of the gravitational field resulting from Riemann curvature in four-dimensional spacetime \cite{5}, gives a microsopic quantum mechanical description of curvature.  For example, in string theory, the curvature of spacetime is physically equivalent to a coherent state (condensate) of closed strings \cite{6}.

In mathematics, category theory, originally defined by Mac Lane and Eilenberg in 1945 \cite{7}, is a unifying language for modern mathematics that can serve as an alternative to set theory, as well provide order for the abstract, large scale  structures that emerge from set theoretic foundations.  The Langlands program \cite{8,9}, an ambitious series of conjectures that relate number theory to harmonic analysis and algebraic geometry, in its most general form, is formulated in the language of category theory.  The proof of Fermat's last theorem \cite{10}, stems from the proof of a special case of the Langlands program, the Taniyama-Shimura conjecture \cite{11}, which states that every elliptic curve is really a modular form in disguise.  In a more general setting \cite{12}, such as the geometric Langlands, one begins by starting with an object that (very abstractly) lies in a category of motives.

Motives were introduced in 1964 by Grothendieck \cite{13}, to express the idea that there is a single universal cohomology theory underlying the various cohomology theories for algebraic varieties.  Motives come in two forms, pure or mixed \cite{14}, depending on the type of algebraic variety attached to the motive.  Mixed motives have found wide mathematical applications \cite{15} in diverse areas from Hodge theory, moduli spaces and algebraic K-theory to polylogarithms, automorphic forms and L-functions.  In a very real sense, the theory of motives is a unifying tool in mathematics that is also finding its way into theoretical physics through scattering amplitudes in twistor space, black hole entropy, and M-theory compactifications.  In this essay, we explore the hypothesis that motives provide a vital, and unreasonably effective ingredient in formulating a unified theory of physics and mathematics.

\section{Motivic Amplitudes}
The concept of a \emph{twistor} was proposed by Penrose in 1967, as a higher dimensional theory of quantum gravity that complexifies the usual 3+1 Minkowski coordinates to a four-dimensional complex space, twistor space \cite{16}.  Research in twistory theory proceeded independently of other approaches to quantum gravity until 2003 when Witten proposed a stringy formulation of super Yang-Mills (SYM) scattering amplitudes in twistor space \cite{17}.  Witten asserted that maximally helicity violating (MHV) amplitudes in twistor space localize on holomorphic curves.  The simplest of these curves are Riemann spheres, which are identified with D-instantons in a topological string theory with super Calabi-Yau twistor target space.

Witten's paper sparked a flurry of follow-up papers, leading up to Arkani-Hamed and Trnka's conclusion \cite{18} that scattering amplitudes in gauge theories have a surprising simplicity and hidden infinite dimensional symmetry, completely obscured in the usual Feynman diagram approach to quantum field theory.  They posited the existence of a new mathematical object, the \emph{amplituhedron}, that generalizes the combinatorial object, the associahedron, for planar Yang-Mills scattering amplitudes.  In this framework, locality and unitary are emergent features arising from positivity of the Grassmannian scattering geometry.  Goncharov et al., building on joint work with Arkani-Hamed, introduced the notion of a \emph{motivic amplitude} to describe planar SYM scattering amplitudes in a completely canonical way \cite{19}, and elucidated their relation to associahedra.  Motivic amplitudes provide a hint that mixed motives may underlie SYM and quantum field theory in general.

\section{Modular Black Hole Entropy}
Three months before his death in 1920, Ramanujan described \emph{mock theta functions} \cite{20} in a letter to Hardy.  Such functions remained obscure until in 2002 it was shown by Zwegers they are incomplete Maass forms and belong to the theory of modular forms \cite{21}, with applications to elliptic curves, singular moduli and Galois representations.

In the study of quantum black holes in string theory, the physical problem of
counting the dimensions of certain eigenspaces (corresponding to the number of 1/4-BPS dyonic states of a
given charge) leads to the study of Fourier coefficients of certain meromorphic Siegel modular
forms and their corresponding generating functions.  These generating functions were found by Dabholkar et al. in 2014 to belong to the class of functions of mock modular forms \cite{22}, which count quantum degeneracies of the black hole horizon.  In related work, Cheng et al. used mock modular forms to introduce \emph{Umbral Moonshine} \cite{23}, an analog of Conway's Monstrous Moonshine, where mock theta functions appear as McKay-Thompson series in the correspondence.

Weight one mock modular forms can be attached to Galois group representations \cite{24}, and thus can be set in the context of the more general Langlands program.  The Langlands program so far has not been concretely connected to Monstrous Moonshine, and it is exciting mock modular forms provide a bridge between the two areas of study.  Recent studies of the Leech lattice and higher dimensional lattices in (bosonic) M-theory \cite{25,26} provide hope that there exists a general motivic framework that will elucidate connections between sporadic groups, unimodular lattices, and quantum black holes.

\section{M-theory Compactifications}
After Witten introduced M-theory in 1995, as a unifying theory in eleven dimensions that has the ten dimensional string theories and D=11 supergravity as its low energy limits \cite{27}, it was noticed that branes are the primary objects in this approach to quantum gravity.  M-theory contains 2-brane and 5-brane soliton solutions, which can arrange themselves in various configurations and appear as tiny quantum black holes when a portion of the eleven dimensions are curled up tightly in $n$-dimensional versions of the torus \cite{28,29}.  When the torus is five-dimensional or higher, exceptional (U-duality) group symmetries arise and the microscopic black holes charges configure in a moduli space that is a hermitian symmetric domain \cite{30}.  In a fully nonperturbative M-theory, the black holes charges are quantized, hence are integer valued and the hermitian symmetric domains take on integral forms.

Hermitian symmetric domains over the integers fall into the realm of Shimura varieties \cite{31}, which are higher dimensional versions of elliptic curves.  Thus, toroidal M-theory compactifications can serve as fertile ground \cite{32} for the study of the Langlands program, in a setting that generalizes the elliptic curve-modular form correspondence used to prove Fermat's last theorem.  Unfortunately, the mathematics of \emph{exceptional Shimura varieties} and their motives remains incomplete, so a physicist or mathematician must tread courageously into this new territory.  The benefits are very rich, however, as such a journey could lead to complete theory of quantum gravity, as well as a unifying theory of mathematics.

\section{Conclusion}
In this essay, it was argued that Grothendieck's mathematical creation, motives, may serve as an unreasonably effective tool leading to a unified theory of physics and mathematics.  Strong evidence for this hypothesis was given in physics, from the study of motivic amplitudes in twistor-string theory, where the very foundations of quantum field theory are currently being re-written in a manifestly non-local and non-unitary way.  Further support for motives was shown to arise in the study of quantum black holes in string theory, where calculations of their entropy reveal a deep connection between number theory and geometry.  Moreover, in the parent M-theory, the black hole configuration spaces, themselves, point to a higher dimensional relationship that may allow the emergence of spacetime itself from motives.

The language of the Langlands program and motives is very abstract, and it is difficult for any single physicist or mathematician to grasp the entire structure.  To that end, it is in the best interest of both disciplines to encourage frequent collaborations, where the intuition provided by physics can assist in the progress of the required mathematics for the deep problem of completing a theory of quantum gravity.  Along the way, surely many unforeseen relationships in mathematics will be illuminated and an elegant tapestry of mathematical relationships will be constructed, that from the viewpoint of many mathematicians, will be indistinguishable from a unified theory of mathematics.  The ancient dream of the Pythagoreans may finally be vindicated via a motivic apeiron.

\end{document}